\documentclass[letterpage,titlepage,11pt]{article}
\usepackage{amsmath,amssymb,amsfonts,color,epsf}    
\usepackage{amssymb,slashed,latexsym,color}
\usepackage{graphics}
\usepackage{cite}
\usepackage[font=footnotesize,labelsep=newline,labelfont=sc,justification=centering,position=top]{caption}
\usepackage{mathtools}

\usepackage{color}
\usepackage[
      colorlinks=true,
      linkcolor=blue,
      urlcolor=blue,
      filecolor=black,
      citecolor=green,
      pdfstartview=FitV,
      linktocpage,
      pdftitle={},
        pdfauthor={x, x, x},
        pdfsubject={},
        pdfkeywords={},
        pdfpagemode=None,
        bookmarksopen=true
      ]{hyperref}
      
\makeatletter
\@addtoreset{equation}{section}
\makeatother

\newcommand{\dif}{\mbox{d}}

\begin{document}

\begin{titlepage}
\begin{flushright}
\hfill{ ICCUB 16-030}
\end{flushright} 
\vskip -1cm

\leftline{}
\vskip 2cm
\begin{center}
{\LARGE \bf Extended Galilean symmetries of non-relativistic  strings%
}

\end{center}
\vskip 1.2cm
\centerline{\large\bf Carles Batlle$^{a}$, Joaquim Gomis$^{b}$ and Daniel Not$^{b}$}
\vskip 0.5cm
\centerline{\sl $^{a}$Departament de Matem\`atiques and IOC,}
\centerline{\sl Universitat Polit\`ecnica de Catalunya}
\centerline{\sl EPSEVG, Av. V. Balaguer 1, E-08808 Vilanova i la Geltr\'u, Spain}
\centerline{\sl $^{b}$Departament de F\'{\i}sica Qu\`antica i Astrof\'{\i}sica}
\centerline{\sl and Institut de Ci\`encies del Cosmos (ICCUB),}
\centerline{\sl Universitat de Barcelona, Mart\'i i Franqu\`es 1, E-08028 Barcelona, Spain}
\smallskip
\vskip 0.5cm

\vskip 1.2cm
\centerline{\large\bf Abstract} \vskip 0.4cm 
\noindent  We consider  two non-relativistic strings and their Galilean symmetries. 
	These strings are obtained as
	the two possible non-relativistic (NR) limits of a relativistic string.
	One of them is  non-vibrating  and represents a continuum of non-relativistic massless particles, and the other one is a
	non-relativistic vibrating string.
	For both cases we write the generator of the most general point transformation and impose the condition  of Noether symmetry.  As a result we obtain  two sets of non-relativistic Killing equations for the vector fields that generate the symmetry transformations. Solving these equations shows that  NR strings exhibit two   extended, 
	infinite dimensional space-time symmetries which contain, as a subset,  the Galilean  symmetries.  For each case, we compute the associated conserved charges and discuss the existence of non-central extensions.

\end{titlepage}
 
\pagestyle{plain}

\section{Introduction}
Holography in string theory  \cite{Maldacena:1997re} \cite{Gubser:1998bc} 
\cite{Witten:1998qj} allows to study properties of relativistic strongly coupled quantum field theories on the screen 
in terms of Einstein classical gravity in the bulk, if the curvature is small. For the case of large  curvature  one should replace classical gravity  by full string theory in the bulk. 
{ In particular in \cite{Berenstein:2002jq}  IIB (super) strings in null geodesics of $AdS_5\times S^5$ have been considered. In this sector  string theory is soluble and the corresponding gauge dual sector has been identified.
}

Non-relativistic holography, the application of holographic ideas to non-relativistic physics, has also been considered.
Non-relativistic holography is useful to study properties of strongly correlated systems in condensed matter (see  \cite{sachdev} \cite{Zaanen:2015oix}
and references therein). In this case one could consider relativistic metrics in the bulk with non-relativistic isometries  \cite{Son:2008ye} 
\cite{Balasubramanian:2008dm} or to consider non-relativistic gravity theories in the bulk; see for example \cite{Janiszewski:2012nb} \cite{Son:2013rqa} \cite{Wu:2014dha}. 
Like in the relativistic case, if the curvature is large one should 
use non-relativistic strings in the bulk.  Vibrating non-relativistic strings were introduced in 
\cite{Gomis:2000bd} \cite{Danielsson:2000gi}. Their action was obtained by the
non-relativistic ``stringy" limit of a relativistic string where not only the time direction is large but also a spatial direction along the string \cite{Gomis:2004pw}. The action was also obtained 
using the non-linear realization technique applied to a ``stringy" contraction of the
Poincare group, called  ``stringy" Galilei algebra  \cite{Brugues:2004an} 
\cite{Brugues:2006nh}. This algebra has two type of non-central extensions. 
{ The ``stringy" limit of (super) strings in $AdS_5\times S^5$ was constructed as another soluble sector of the AdS/CFT correspondence \cite{Gomis:2005pg}.
}

There is also the ordinary 
non-relativistic limit of the relativistic string where only the time direction becomes large;  in this case  the non-relativistic string 
does not vibrate \cite{Yastremiz:1991jp}.  
It represents a continuum of non-relativistic massless particles \cite{Duval:2009vt} with an energy density which depends on the position of the particle in the string.
The action is invariant under Galilean transformations that close  without central extension.
The Lagrangian  can be obtained also by the method of  non-linear realizations. 

The first check that holographic ideas are working  is to verify that the symmetries of the bulk and the screen coincide.  In the case of non-relativistic holography one should check that the Galilean symmetries in the bulk and in the screen coincide.

As a first step towards using non-relativistic strings in non-relativistic
holography, we  study in this paper the general Noether Galilean symmetries of  non-relativistic strings.
We will write the generator of the most general point transformation and will impose on it  the condition of Noether symmetry, which  gives the associated non-relativistic Killing equations that one must solve. 
For a massive non-relativistic particle,  the maximal set of 
symmetries is larger than the  Galilei group,  and it is in fact the Schr\"odinger group
\cite{Niederer:1972zz}
\cite{Hagen:1972pd}, which is the group corresponding to the
$z=2$ case of an infinite set of $z$-Galilean conformal algebras \cite{Henkel:1997zz}\cite{negro1}\cite{negro2}.

For a relativistic string, one can consider two non-relativistic (NR) limits according to whether one or two longitudinal coordinates are scaled, which we call, respectively NR particle limit ans NR stringy limit.
For the particle limit, a non-vibrating non-relativistic string is obtained, whose maximal set of symmetries turns out to be
an infinite dimensional
extension of the Galilean conformal algebra  \cite{Henkel:1997zz}\cite{Duval:2009vt}\cite{Hosseiny:2009jj}\cite{Martelli:2009uc}, but with dynamical exponent $z=-1$.   { The enhancement of non-relativistic symmetries for non-relativistic gauge field theories has also been discussed in \cite{Festuccia:2016caf}\cite{Bagchi:2015qcw}\cite{Bagchi:2014ysa}. However, the extended algebras that we obtain in the NR particle limit seem to be fundamentally different due to their stringy origin, and this is reflected, in particular, in the negative values of the dynamical exponent.}

The particle limit of (super) strings in $AdS_5\times S^5$  would lead also  to a soluble sector 
of AdsS/CFT correspondence and is therefore worth to explore.

For the stringy limit, a non-relativistic vibrating string is obtained, with a  new infinite dimensional  algebra, which we call the stringy  Galilean conformal algebra. The  conserved charges are computed and we show the existence of an infinite set of  non-central extensions. 

For general a $p$-brane, the $p+1$ longitudinal directions allow to consider $p+1$ different non-relativistic limits. In this paper we study only the case of the particle limit,  which turns out to have a group of transformations corresponding to an infinite dimensional  extension of the Galilean algebra with $z=-p$.

The organization of the paper is as follows.  In Section \ref{NRL} we  discuss the conditions under which the terms that are obtained in the expansion of the relativistic action are invariant under Galilean transformations, and review the construction of the two non-relativistic limits for a relativistic string. The non-relativistic particle limit of a general $p$-brane is discussed as well.
In Section \ref{EOM} we write down the equations of motion for both non-relativistic strings and, in particular, for the non-vibrating string, we discuss the gauge fixing procedure and the resulting dynamics.

In Section \ref{particlelimit} we consider  the most general generator of point transformations for the non-vibrating string. The Noether condition for symmetries is applied and a set of 
non-relativistic Killing equations for the symmetry generators are obtained. After solving them, we obtain the algebra of symmetry transformations. Results are also presented for the NR particle limit of a $p$-brane. 

In Section \ref{stringlimit} we apply the same procedure to the NR stringy limit of the string and discuss the corresponding algebra of transformations and the appearance of non-central extensions.   Finally, we review our results and consider some open problems in Section \ref{conclusions}.

\section{Non-relativistic limits of a relativistic extended object \label{NRL}}

In this section we will study the possible non-relativistic limits of an extended object, a
relativistic p-brane. We first analyze some general properties on the non-relativistic limit.  Consider a general action of a relativistic extended object with coordinates $X$'s\footnote{The results obtained will also apply to any relativistic field theory Lagrangian.},

\begin{equation}\label{NRlim1}
	S[X] = \int {\cal L}[X]
\end{equation}	
we assume that the Lagrangian density ${\cal L}$ is pseudo-invariant under a given set of   relativistic symmetries $\delta_R$,
\begin{equation}\label{NRlimit2}
	\delta_R {\cal L} = \partial\cdot F,	
\end{equation}	
where $\partial\cdot F$ denotes the divergence of $F$.
In order to study the non-relativistic limit we introduce a dimensionless parameter $\omega$ 
and we scale the variables and  constants of the Lagrangian and relativistic transformations accordingly.

We assume that the Lagrangian density and the symmetry transformation can be expanded in powers of 
$\omega$,\footnote{The details of the expansion may depend on the system and symmetry that one is considering,  but the same ideas can be applied to other cases. {  This particular expansion is motivated by the result that one obtains in the case of the NR limit of a relativistic particle, and is also useful for the sytems that we are considering.}}
\begin{eqnarray}
	\delta_R &=& \delta_0 + \omega^{-2} \delta_{-2}+\dots,\label{NRlimit3}\\
	{\cal L} &=& \omega^2 {\cal L}_2 + {\cal L}_0 + \omega^{-2}{\cal L}_{-2}+\dots,
	\label{NRlimit4}\\
	F &=& \omega^2 F_2 + F_0 + \omega^{-2} F_{-2}+\dots.\label{NRlimit5}	
\end{eqnarray}	
where the first term in the expansion of the relativistic symmetry, 
$\delta_R$,  is the non-relativistic transformation $\delta_0$.\footnote{The analysis below can also be used  for the Carroll limit that was introduced as a different type of non-relativistic limit by Levy-Leblond \cite{Levy-Leblond};  see also \cite {dubouis}.}
Condition (\ref{NRlimit2}) implies and infinite set of conditions. For the first orders in the expansion parameter we have
\begin{eqnarray}
	\delta_0 {\cal L}_2 &=& \partial\cdot F_2,\label{NRlimit6}\\
	\delta_0 {\cal L}_0 + \delta_{-2} {\cal L}_2 &=& \partial\cdot F_0,\label{NRlimit7}\\
	\delta_0 {\cal L}_2 + \delta_{-2} {\cal L}_0 + \delta_{-4}{\cal L}_2 &=& \partial\cdot F_{-2}.\label{NRlimit8}
\end{eqnarray}	
From these one sees that the highest  order Lagrangian density ${\cal L}_2$, which appears with a divergent factor $\omega^2$,  is always (pseudo)invariant under $\delta_0$, while the finite one, ${\cal L}_0$ is only invariant if $\delta_{-2}{\cal L}_2$ is  a divergence. The latter condition happens, in particular, if ${\cal L}_2$ itself is itself a divergence or, even more particularly, if ${\cal L}_2=0$, which will one of the cases   that we will be considering.
In the limit $\omega\to\infty$ all the terms with negative powers of $\omega$ 
vanish.

Let us apply these ideas to
a relativistic string in flat space-time described by
the Nambu-Goto action
\begin{equation}\label{actionNG}
	S=-T\int\mbox{d}^2\sigma\ \sqrt{-\det G},
\end{equation}
where  $G$ is the induced metric, with elements 
\begin{equation}\label{ind_metric}
	G_{\alpha\beta}=\partial_{\alpha}X^{\mu}\partial_{\beta}X_{\mu},
\end{equation}
and where  $X^\mu(\tau,\sigma)$, $\mu=0,1,\ldots d$ are the embedding coordinates in flat $(d+1)$-spacetime, while the world-volume coordinates are $\sigma^\alpha=(\sigma^0,\sigma^1)=(\tau,\sigma)$
$\alpha,\beta=0,1$. As usual, we will denote derivatives with respect to $\tau$ and $\sigma$ by dots and primes, respectively.

For a string, one can consider two non-relativistic limits, which depend on whether one or two embedding coordinates are scaled by $\omega$.  
The \textit{particle limit} is obtained for $X^0=\omega t$, $X^i=x^i$, $i=1,\ldots,d$, for which \cite{Yastremiz:1991jp}
\begin{equation}
	S_{NG} = -T \omega \int \dif\tau\dif\sigma\	 \sqrt{(\dot t \vec{x}\,' - t' \dot{\vec{x}})^2} +
	O(\frac 1{\omega})
	\label{LPL}
\end{equation}
After re-scaling  $T\omega =T_{NR}\equiv T$, the first term in the expansion is a finite one in $\omega$,
\begin{equation}\label{LPLf}
	S_{PL} = -T  \int \dif\tau\dif\sigma\	 \sqrt{(\dot t \vec{x}\,' - t' \dot{\vec{x}})^2},
\end{equation}
and, according to the general discussion, the Lagrangian density $   -T   \sqrt{(\dot t \vec{x}\,' - t' \dot{\vec{x}})^2}      $ will be, at least, pseudo-invariant under  Galilean transformations.  
In fact, the Galilean symmetry transformations obtained by the NR limit 
\begin{equation}
	\delta t=\beta,
	\quad 
	\delta x_i=a_i + b_i t + \omega_{ij}x_j,\quad \omega_{ij}=-\omega_{ji},
\end{equation}
close under the Galilei algebra without central extension \textit{i.e.}, the Lagrangian is invariant under the Galilean transformations and not pseudo invariant \cite{levyleblond69} \cite{Marmo:1987rv}.  
{  From  (\ref{LPLf}) it follows also that the action is invariant under the scaling
\begin{equation}
t \to \lambda t,\quad \vec{x}\to \lambda^{-1} \vec x,
\end{equation}
which indicates a negative dynamical exponent. This will be reflected in the full extended Galilean transformations that we will construct.
}

The \textit{stringy limit} is obtained with $X^L=\omega \hat X^L $, $L=0,1$, $X^i=y^i$, $i=2,\ldots,d$\cite{Gomis:2000bd} \cite{Danielsson:2000gi} (see also \cite{Gomis:2004pw}).
After some algebra, one obtains
\begin{equation}\label{NRlimit16}
	\sqrt{-\det G}=w^{2}\sqrt{-\det\hat{G}}+\frac{1}{2}\sqrt{-\det\hat{G}}\left(\hat{G}^{-1}\right)^{\gamma\delta}\tilde{G}_{\gamma\delta}+\mathcal{O}\left(\frac{1}{w^{2}}\right),
\end{equation}
where
\begin{eqnarray}
	\hat{G}_{\alpha\beta} &=& 	\partial_\alpha \hat{X}^L\partial_\beta \hat{X}_L,\\
	\tilde{G}_{\alpha\beta} &=& 	\partial_\alpha y_i\partial_\beta y_i.
\end{eqnarray}	
The term $\sqrt{-\det\hat{G}}$ is, in fact,  a total derivative,  since
\begin{equation}\label{totalder1}
	\det\hat{G} = - \left[      \partial_\tau (X^0 \partial_\sigma X^1)+ \partial_\sigma (-X^0\partial_\tau X^1) \right]^2,
\end{equation}
and hence, according to the previous discussion, the action \cite{Gomis:2004pw}
\begin{equation}\label{slaction}
	S_{SL}=-\frac{T}{2}\int\mbox{d}\tau\mbox{d}\sigma\ \sqrt{-\det\hat{G}}\left(\hat{G}^{-1}\right)^{\gamma\delta}\tilde{G}_{\gamma\delta}
\end{equation}
will be invariant under Galilean transformations
\begin{equation}\label{transfor}
	\delta\hat X^L=\epsilon^L+ \omega^L{}_M\hat X^M,\quad \delta y^i=
	\epsilon^i+ \omega^i{}_j y^j+\omega^i{}_L\hat X^L.
\end{equation}

After some algebra, the corresponding Lagrangian density can be written as
\cite{Garcia:2002fa}
\begin{equation}\label{LSL}
	{\cal L}_{SL}=\frac{T}{2}\frac{1}{\left(\dot{t}z'-t'\dot{z}\right)}\left[\left(-t'^{2}+z'^{2}\right)\dot{\vec{y}}^{\, 2}+\left(-\dot{t}^{2}+\dot{z}^{2}\right)\vec{y}\, '^{\, 2}+2\left(t'\dot{t}-z'\dot{z}\right)\dot{\vec{y}}\cdot\vec{y}\,'\right],
\end{equation}
with  $\hat{X}^0=t$, $\hat{X}^1=z$.

For a $p$-brane it makes sense to consider $p+1$ different non-relativistic limits, according to the number ($1,2,\ldots, p+1$) of embedding coordinates that are scaled.  The case of the NR $p$-brane limit of a $p$-brane, which for $p=1$ corresponds to the NR stringy limit of the string, was discussed in \cite{Brugues:2006nh}.

	Here we
consider the NR particle limit of a $p$-brane in flat spacetime, for general $p$. The relativistic Lagrangian density is given by
\begin{equation}\label{pbrane1}
	{\cal L}=-T \sqrt{-\det G}, \quad G_{\alpha\beta} = \partial_\alpha X^\mu \partial_\beta X_\mu,\quad \alpha,\beta=0,1,\ldots,p+1,\ \ \mu=0,1,\ldots,d.	
\end{equation} 
As in the string case, the NR particle limit is obtained with $X^0=\omega t$, $X^i=x^i$, $i=1,2,\ldots,d$. One has
\begin{equation}\label{pbrane2}
	G_{\alpha\beta} = -\omega^2 \partial_\alpha t \partial_\beta t + \tilde{G}_{\alpha\beta},\quad\text{with}\ \tilde{G}_{\alpha\beta} = \partial_\alpha x_i \partial_\beta x_i.	
\end{equation} 
Then
\begin{equation}\label{pbrane3}
	\begin{split}
		\det G = & \frac{1}{(p+1)!} \epsilon^{\alpha_1\alpha_2\ldots\alpha_{p+1}}	\epsilon^{\beta_1\beta_2\ldots\beta_{p+1}} G_{\alpha_1\beta_1}G_{\alpha_2\beta_2}\cdots G_{\alpha_{p+1}\beta_{p+1}}\\
		=& \frac{1}{(p+1)!} \epsilon^{\alpha_1\alpha_2\ldots\alpha_{p+1}} \epsilon^{\beta_1\beta_2\ldots\beta_{p+1}}\\
		&(-\omega^2 \partial_{\alpha_1}t \partial_{\beta_1}t + \tilde{G}_{\alpha_1\beta_1})\\
		& (-\omega^2 \partial_{\alpha_2}t \partial_{\beta_2}t + \tilde{G}_{\alpha_2\beta_2})\\
		& \vdots \\
		&(-\omega^2 \partial_{\alpha_{p+1}}t \partial_{\beta_{p+1}}t + \tilde{G}_{\alpha_{p+1}\beta_{p+1}}).
	\end{split}
\end{equation}	 
When forming the products, any term with a power higher than $\omega^2$ will have at least two derivatives of $t$ with indexes belonging to the same $\epsilon$, and hence will cancel by symmetry. The only surviving terms are the $p+1$ ones with power $\omega^2$ (which are all equal due to symmetry), and the term with  $p+1$ factors of elements of $\tilde{G}$, which is equal to $\det\tilde{G}$,
\begin{equation}\label{pbrane4}
	\det G = -\omega^2 \frac{1}{p!} \	\epsilon^{\alpha_1\alpha_2\ldots\alpha_{p+1}}\	\epsilon^{\beta_1\beta_2\ldots\beta_{p+1}}\ \partial_{\alpha_1}t\ \partial_{\beta_1}t\
	\tilde{G}_{\alpha_2\beta_2}\cdots\tilde{G}_{\alpha_{p+1}\beta_{p+1}} + \det\tilde{G},
\end{equation}	  
The expansion in $\omega$ of the relativistic Lagrangian density is 
\begin{equation}\label{pbrane5}
	{\cal L} = -T\omega \sqrt{	
		\frac{1}{p!} \	\epsilon^{\alpha_1\alpha_2\ldots\alpha_{p+1}}\	\epsilon^{\beta_1\beta_2\ldots\beta_{p+1}}\ \partial_{\alpha_1}t\ \partial_{\beta_1}t\
		\tilde{G}_{\alpha_2\beta_2}\cdots\tilde{G}_{\alpha_{p+1}\beta_{p+1}}	
	}
	+ \mathcal{O}\left(\frac{1}{\omega}\right).
\end{equation}
According to the general discussion above, the action with Lagrangian density
\begin{equation}\label{pbrane6}
	{\cal L}_{PL} = -\hat T \sqrt{	
		\frac{1}{p!} \	\epsilon^{\alpha_1\alpha_2\ldots\alpha_{p+1}}\	\epsilon^{\beta_1\beta_2\ldots\beta_{p+1}}\ \partial_{\alpha_1}t\ \partial_{\beta_1}t\
		\tilde{G}_{\alpha_2\beta_2}\cdots\tilde{G}_{\alpha_{p+1}\beta_{p+1}}	
	},
\end{equation}
where $\hat T=\omega T$, will be NR (pseudo)invariant. For $p=1$ one recovers the  action (\ref{LPLf}) for the NR string, while $p=0$ yields a particle Lagrangian which is just a total derivative $-\hat T \dot t$. In this case, the next term in the $\omega$ expansion will also be NR invariant, and it results in the standard NR particle action.

\section{Equations of motion of non-relativistic strings}\label{EOM}
The Lagrangian densities for the extended NR objects that we have constructed in the previous section are homogeneous functions of first order of the derivatives of the field variables (and in particular of the derivatives with respect to $\tau$), and hence  are invariant under diffeomorphisms, have an identically zero canonical Hamiltonian, and yield $p+1$ first-class constraints $\Phi_a(\vec x,t,\vec p,E)=0$, $a=0,\ldots,p$. The canonical action is thus given by
\begin{equation}
	\label{killingCBA4}
	S=\int \dif\tau\dif\sigma^1\cdots \dif\sigma^p \left( \dot x_i p_i -\dot t E  - \sum_{a=0}^p \lambda_a \Phi_a(\vec x,t,\vec p,E)
	\right),
\end{equation}
with $\{\lambda_a\}$ a set of $p+1$ Lagrange multipliers. 


Let us consider  first the Lagrangian density corresponding to the action (\ref{LPL}),
\begin{equation}\label{NRPLs1}
	{\cal L} = -T 	\sqrt{(\dot t \vec{x}\,' - t' \dot{\vec{x}})^2} = -T \sqrt{\vec B^2}, \quad \vec B = 	\dot t \vec{x}\,' - t' \dot{\vec{x}}.
\end{equation}
The canonical momenta are
\begin{eqnarray}
	E &=& -	\frac{\partial {\cal L}}{\partial \dot t} = \frac{T}{\sqrt{{\vec{B}}^2}} \vec B\cdot \vec{x}',\label{NRPLs2}\\
	\vec p &=& \frac{\partial{\cal L}}{\partial \dot{\vec{x}}} = \frac{T}{\sqrt{\vec B^2}} t' \vec B,\label{NRPLs3}	
\end{eqnarray}	
and yield the primary first-class constraints
\begin{eqnarray}
	\Phi_0 &=&\frac{1}{2T} \left(\vec{p}\ ^2 - T^2 (t')^2\right),\label{NRPLs4}\\
	\Phi_1 &=& \vec p\cdot\vec{x}' - E t',\label{NRPLs5}	
\end{eqnarray}	
with equal-$\tau$ Poisson brackets  
\begin{equation}
	\begin{split}
		\left\{ \Phi_{0}\left(\sigma\right),\Phi_{0}\left(\tilde{\sigma}\right)\right\} &= \Phi_{1}\left(\sigma\right)\partial_{\sigma}\delta\left(\sigma-\tilde{\sigma}\right)-\Phi_{1}\left(\tilde{\sigma}\right)\partial_{\tilde{\sigma}}\delta\left(\sigma-\tilde{\sigma}\right),\\
		\left\{ \Phi_{1}\left(\sigma\right),\Phi_{0}\left(\tilde{\sigma}\right)\right\} &= \Phi_{0}\left(\sigma\right)\partial_{\sigma}\delta\left(\sigma-\tilde{\sigma}\right)-\Phi_{0}\left(\tilde{\sigma}\right)\partial_{\tilde{\sigma}}\delta\left(\sigma-\tilde{\sigma}\right), \\
		\left\{ \Phi_{1}\left(\sigma\right),\Phi_{1}\left(\tilde{\sigma}\right)\right\} &= \Phi_{1}\left(\sigma\right)\partial_{\sigma}\left(\delta\left(\sigma-\tilde{\sigma}\right)\right)-\Phi_{1}\left(\tilde{\sigma}\right)\partial_{\tilde{\sigma}}\delta\left(\sigma-\tilde{\sigma}\right).
	\end{split}\label{PL-PB}
\end{equation}

The canonical action is 	
\begin{equation}\label{NRPLs6}
	S=\int\text{d}\tau\text{d}\sigma\left(-E\dot{t}+\vec{p}\cdot\dot{\vec{x}}-\frac{1}{2T}\lambda(\vec{p}\ ^{2}-T^{2}{t'}^{2})-\mu(\vec{p}\cdot\vec{x}'-E t')\right),
\end{equation}
and the corresponding equations of motion are
\begin{eqnarray}
	\dot{E}	&=&	\left(T\lambda t'+\mu E\right)',\label{PLdyn1} \\
	\dot{\vec{p}}	&=&	\left(\mu\vec{p}\right)',\label{PLdyn2} \\
	\dot{t}	&=&	\mu t',\label{PLdyn3} \\
	\dot{\vec{x}}	&=&	\frac{\lambda}{T}\vec{p}+\mu\vec{x}'\label{PLdyn4},
\end{eqnarray}
provided that the  terms coming from partial integrations in $\sigma$ are zero.

For closed strings  this is automatic, while for open strings one has to demand that
\begin{eqnarray}
	\left. \left(\left(   T \lambda t' + \mu E         \right)\delta t- \mu p_i \delta x_i\right) \right|_B &=&0,\label{plBC1}
\end{eqnarray}	
where $_B$ indicates evaluation at the boundary of $\sigma$. In order to satisfy this one can impose\footnote{We thank Paul Townsend for remarks concerning the boundary conditions.}
\begin{eqnarray}
	\left. \mu\right|_B &=& 0,\label{plBC1b}\\	
	\left. t' \delta t\right|_B  &=& 0.\label{plBC2b}
\end{eqnarray}	
Notice that (\ref{plBC2b}) is satisfied for both Dirichlet and Neumann conditions on $t$. Conservation of total energy, however, requires Neumann conditions (see (\ref{timeH}) below) and hence we will take as our boundary conditions
\begin{eqnarray}
	\left. \mu\right|_B &=& 0,\label{plBC2}\\	
	\left. t'\right|_B  &=& 0.\label{plBC3}
\end{eqnarray}

One can compute the canonical generator of the gauge transformations  
and get the following transformations of the state variables,
\begin{eqnarray}
	\delta_{\text{Diff}} x_i &=& \frac{1}{T} \epsilon_0 p_i + \epsilon_1 x_i',\\
	\delta_{\text{Diff}} t &=& \epsilon_1 t',
\end{eqnarray}
with $\epsilon_0$ and $\epsilon_1$ arbitrary functions of $\tau$, $\sigma$, and with $E$ and $p_i$  given by 
(\ref{NRPLs2}) and  (\ref{NRPLs3}), respectively.

One can check that under these transformations the Lagrangian density (\ref{NRPLs1}) varies as
\begin{equation}
	\delta_{\text{Diff}}{\cal L} = \partial_\tau\left(   T\epsilon_0 (t'^2   -\dot t t')      \right) + \partial_\sigma\left( \epsilon_1 {\cal L}                \right).	
\end{equation}

Equations (\ref{PLdyn1})---(\ref{PLdyn4}) contain two Lagrangian multipliers $\lambda(\tau,\sigma)$ and $\mu(\tau,\sigma)$. If we consider the conformal gauge $\lambda =1, \mu =0$ the equations of motion become
\begin{eqnarray}
	\dot{E}	&=&	T t^{''},\label{PLdyn1a} \\
	\dot{\vec{p}}	&=&	0,\label{PLdyn2a} \\
	\dot{t}	&=& 0,\label{PLdyn3a} \\
	\dot{\vec{x}}	&=&\frac{1}{T}\vec{p}\label{PLdyn4a},
\end{eqnarray}
and they  should be supplemented with the constraints  (\ref{NRPLs4}), (\ref{NRPLs5}).

The equations of motion for any value of $\sigma$ are the  ones of a massless Galilean 
particle  \cite{Duval:2009vt}, except for the fact that now we have (\ref{PLdyn1a}) instead of $\dot E=0$. 
The total energy
\begin{equation}
	H(\tau) = \int\dif\sigma\  E(\tau,\sigma)
\end{equation}	
is  nevertheless conserved. Indeed
\begin{equation}\label{timeH}
	\dot H = \int\dif\sigma\ \dot E = \int\dif\sigma\ T t'' = T \int\dif\sigma\  \partial_\sigma t', 
\end{equation}
which is zero for closed strings, but also for open strings, due to the boundary condition 	(\ref{plBC3}). It follows from (\ref{PLdyn1a}) that energy flows along the string according to the curvature of the $t$ coordinate, and this is a unique feature of having an extended object instead of a particle.

In order to study the reduced physical space we must
introduce two gauge fixing constraints, 
$\Psi_0$ and $\Psi_1$, such that  $\{\Phi_0,\Phi_1,\Psi_0,\Psi_1\}$ is a set of  second-class constraints on the variety ${\cal M}$ defined by all four constraints. The arbitrary functions can then be determined by demanding the stability of $\Psi_0$ and $\Psi_1$. 

Let us remark first that the naive gauge fixing condition $\Psi(\tau,\sigma)=t(\tau,\sigma)-\frac{1}{\sqrt{T}}\tau$ is not acceptable. Indeed, one has that $\{\Phi_0,\Psi\}=0$ trivially and
\begin{equation}
	\left.  \{\Phi_1(\tau,\sigma) , \Psi(\tau,\tilde\sigma)        \}\right|_{\cal M} = \left. t'(\tau,\sigma)\delta(\sigma-\tilde\sigma)\right|_{\cal M} = 0 	
\end{equation} 
where $t(\tau,\sigma)=\frac{1}{\sqrt{T}}\tau$ on ${\cal M}$ has been used in the last step. Hence, a condition of the form $t\sim\tau$ does not render second class any of the original primary constraints, and does not determine any of the  Lagrangian multipliers $\lambda, \mu$. From the above computation it is clear that an acceptable gauge condition has to include a dependence on $\sigma$ for $t$, and we choose
\begin{equation}\label{Psi0}
	\Psi_0(\tau,\sigma)= t(\tau,\sigma)- f(\sigma),	
\end{equation}	 
where we leave $f$ free so that the boundary conditions for the open string, {  or the winding number assignement for the closed string,} can be satisfied. Since $t$ cannot be chosen as  the evolution parameter of the gauge-fixed system, a sensible  second gauge condition is given by
\begin{equation}\label{Psi1}
	\Psi_1(\tau,\sigma)= x_1(\tau,\sigma)- \frac{1}{T}p_1(\tau,\sigma)\tau.	
\end{equation}	 
One has then the non-zero Poisson brackets
\begin{eqnarray}
	\left.  \{\Phi_1(\tau,\sigma) , \Psi_0(\tau,\tilde\sigma)        \}\right|_{\cal M} &=& f'(\sigma) \delta(\sigma-\tilde\sigma),\\
	\left.  \{\Phi_0(\tau,\sigma) , \Psi_1(\tau,\tilde\sigma)        \}\right|_{\cal M} &=& \left.  -\frac{1}{T} p_1(\tau,\sigma) \delta(\sigma-\tilde\sigma)\right|_{\cal M} =-\frac{1}{T}p_1^*(\tau,\sigma) \delta(\sigma-\tilde\sigma),\\
	\left.  \{\Phi_1(\tau,\sigma) , \Psi_1(\tau,\tilde\sigma)        \}\right|_{\cal M} &=& -\frac{1}{T}\tau p_1^*(\tau,\tilde\sigma)\partial_{\sigma} \delta(\sigma-\tilde\sigma),
\end{eqnarray}	
where $p_1^*$ means $p_1$ expressed in terms of $f'$ and $p_i$, $i=2,\ldots,d$, using  $\Phi_0$,
\begin{equation} 	
	p_1^*(\tau,\sigma)=	\left(  T^2 (f'(\sigma))^2 - \sum_{i=2}^d p_i^2(\tau,\sigma)  \right)^{1/2}.
\end{equation}
The stability of the gauge conditions yields
\begin{eqnarray}
	\left. \dot\Psi_0(\tau,\sigma)\right|_{\cal M} &=& - f'(\sigma) \mu(\tau,\sigma),\\
	\left. \dot\Psi_1(\tau,\sigma)\right|_{\cal M} &=& \frac{1}{T}\lambda(\tau,\sigma) p_1^*(\tau,\sigma)+ \left(\mu(\tau,\sigma)\tau p_1^*(\tau,\sigma)\right)' -\frac{1}{T}p_1^*(\tau,\sigma).
\end{eqnarray}	

Requiring  these to be zero one has $\mu=0$ and $\lambda=1$, and hence (\ref{Psi0}) and (\ref{Psi1}) define  a conformal gauge.
$E$ is expressed in terms of the physical variables using 
$\Phi_1=0$.  The transverse coordinates $x_i(\sigma)$,  $p_i(\sigma)$, $i=2,\ldots, d$,  with equations of motion 
\begin{eqnarray}  
	\dot p_i &=& 0, \\
	\dot x_i &=& \frac{1}{T}p_i,
\end{eqnarray}	
with $x_1$  the evolution parameter, constitute the reduced, or physical, phase space.  From these equations one can see that the physical degrees of freedom do not vibrate, and constitute a continuum of massless Galilean particles, in the sense that the corresponding algebra of conserved charges associated to the Galilean symmetries does not exhibit any central extension. {  It should be noticed \cite{Gomis:2016zur} that the physical degrees of freedom describe a Galilean string with vanishing potential energy density, although the tension parameter $T$ in the NR lagrangian density is finite. On the other hand, non-relativistic and carrollian symmetries of relativistic tensionless strings have been discussed in \cite{Bagchi:2013bga}\cite{Duval:2014lpa}.}

Let us consider now the NR stringy limit of the string.
The Lagrangian density (\ref{LSL})
can be written in terms of light-cone coordinates $r=t-z$, $s=t+z$ as
\begin{equation}\label{Lrs}
	{\cal L}_{rs}= \frac{T}{\dot r s' - r' \dot s}\left(
	-r's' \dot y_i^2 - \dot r \dot s y_i'^2 + (\dot r s' + r' \dot s) \dot y_i  y_i'
	\right).
\end{equation}
The canonical momenta are given by
\begin{eqnarray}
	\pi_r &=& \frac{\partial {\cal L}_{rs}}{\partial \dot r} = -\frac{s'}{\dot r s' - r' \dot s}{\cal L}_{rs} + T \frac{-\dot s y_i'^2 + s' \dot y_i  y_i'}{\dot r s' - r' \dot s} ,\label{slpr}\\
	\pi_s &=& \frac{\partial {\cal L}_{rs}}{\partial \dot s} = \frac{r'}{\dot r s' - r' \dot s}{\cal L}_{rs} + T \frac{-\dot r y_i'^2 + r' \dot y_i  y_i'}{\dot r s' - r' \dot s} ,\label{slps}\\
	p_i &=&   \frac{\partial {\cal L}_{rs}}{\partial \dot y_i}  = \frac{-2 T r's' \dot y_i}{\dot r s' - r' \dot s} + T \frac{\dot r s' + r' \dot s}{\dot r s' - r' \dot s}y_i',\label{slpy}
\end{eqnarray}	
and yield the primary  first class constraints \cite{Gomis:2004ht}
\begin{eqnarray}
	\Phi_0 &=& \frac{1}{2T}\left(p_i^2+{T^2} y_i'^{2}\right)-\pi_ r r'+\pi_ s s',\label{slphi0}\\
	\Phi_1 &=& \pi_r r'+\pi_s s' + p_i y_i',  \label{slphi1} 
\end{eqnarray}
with exactly the same equal-$\tau$ Poisson brackets (\ref{PL-PB}) of the non-vibrating case.

The canonical action is 
\begin{equation}
	S=\int\dif \tau\dif\sigma\left(\pi_r \dot r + \pi_s\dot s +p_i \dot y_i-\lambda\Phi_0-\mu\Phi_1\right),
\end{equation}
from which one can compute the  equations of motion for the fields,
\begin{equation}
	\begin{split}
		\dot{r} &= - (\lambda-\mu)r', \\
		\dot{s} &= (\lambda+\mu)s', \\
		\dot{y_i} &= \lambda\frac{1}{T}p_i+\mu y_i', \\
		\dot{\pi_r} &= -((\lambda-\mu)\pi_r)', \\
		\dot{\pi_s} &= ((\lambda+\mu)\pi_s)', \\
		{\dot{p}_i} &= (T\lambda y_i' + \mu p_i)'.
	\end{split}
\end{equation}

In order to get these equations, one has to cancel the boundary terms coming from integrations in $\sigma$, namely
\begin{equation}
	\label{slBT}
	\left.
	-\lambda \left(
	T y_i' \delta y_i - \pi_r \delta r + \pi_r \delta s
	\right)
	-\mu \left(
	\pi_r \delta r + \pi_s \delta s + p_i \delta y_i
	\right)
	\right|_B=0.
\end{equation}

For closed strings   this is again automatic, while for open strings one can satisfy them by demanding that  
\begin{equation}\label{slBCa}
	\left. \mu\right|_B =0,\quad \left. \pi_r\right|_B =0,\quad \left. \pi_s\right|_B =0,\quad \left. y_i'\right|_B =0.
\end{equation}

From (\ref{slpr}) and (\ref{slps}) it follows then that  $r'=0$ and $s'=0$ also at the boundary, and then, from (\ref{slpy}), that $p_i=0$ at the boundary too. Hence, together with (\ref{slBCa}), we have
\begin{equation}\label{slBCb}
	\left. p_i\right|_B = 0,\quad  \left. r'\right|_B = 0,\quad \left. s'\right|_B = 0.
\end{equation}

After computing the canonical generator of the gauge transformations one obtains for the state variables the following diffeomorphism transformations
\begin{eqnarray}
	\delta_{\text{Diff}} r &=& -(\epsilon_0-\epsilon_1)r',\\
	\delta_{\text{Diff}} s &=& (\epsilon_0+\epsilon_1)s',\\
	\delta_{\text{Diff}} y_i &=&  \frac{1}{T}\epsilon_0 p_i + \epsilon_1 y_i',
\end{eqnarray}
with $\epsilon_0$ and $\epsilon_1$ arbitrary functions of $\tau, \sigma$ and with $p_i$ given by (\ref{slpy}).
Notice that for the covariant longitudinal variables $t, z$ one has
\begin{eqnarray}
	\delta_{\text{Diff}} t &=& \epsilon_0 z' +\epsilon_1 t'  ,\\
	\delta_{\text{Diff}} z &=& \epsilon_0 t' + \epsilon_1 z'. 
\end{eqnarray}

If we consider the conformal gauge $\lambda=1, \mu =0$ the equations of motion become
in \cite{Gomis:2004ht}
\begin{equation}
	\begin{split}
		\dot{r} &= -r', \\
		\dot{s} &= s', \\
		\dot{y_i} &= \frac{1}{T}p_i, \\
		\dot{\pi_r} &=- \pi_r', \\
		\dot{\pi_s} &= \pi_s', \\
		{\dot{p}_i} &= Ty_i'',
	\end{split}
\end{equation}
and we get classical string vibrations in the transverse directions, $\ddot y_i = y_i''$.
We refer to \cite{Gomis:2004ht} for a discussion of the classical dynamics of this system.
At quantum level  it is described by a conformal field theory, with critical dimension $D=26$ (see \cite{Gomis:2000bd} \cite{Danielsson:2000gi}).

\section{Space-time symmetries of the non-vibrating NR string}\label{particlelimit}

In this section we will derive spacetime symmetries for the non-vibrating NR string and also for the corresponding higher dimensional objects obtained from relativistic $p$-branes.
We will construct all the spacetime symmetries 
given by point transformations, that is, transformations generated by generators of the form
\begin{equation}
	\label{killingCBA1}
	G(\tau) = \int \dif\sigma^1\cdots\dif\sigma^p \left(\xi_i(\vec x,t)p_i - \xi_0(\vec x,t)E + F(\vec x,t)\right),
\end{equation}
where we have grouped into $\vec x=(x_i)$ all the space-like variables, $\vec p=(p_i)$ is the canonical momentum associated to $\vec x$ and $E$ is the canonical momentum corresponding to $t$, with the sign convention
$E=-\frac{\partial {\cal L}}{\partial \dot t}$.
The function $F$ is zero if the transformation leaves invariant the Lagrangian density, and different from zero if it is only pseudoinvariant (the variation yields a total derivative)\cite{levyleblond69} \cite{Marmo:1987rv}.
Functions $\vec\xi$, $\xi^0$ and $F$ will be determined by demanding that $G$ be a constant of movement. Noether's theorem for the Hamiltonian formalism ensures then that $G$ generates symmetries of the action.

We consider the following generator of space-time transformations 
\begin{equation}\label{NRPLs7}
	G=\int\text{d}\sigma\left(\vec{\xi}(t,\vec{x})\cdot\vec{p}-\xi_0(t,\vec{x})E\right).
\end{equation}
Since the action is exactly invariant under Galilean transformations, we make the \textit{ansatz} that this will also be the case for the extended transformations that we are looking for, and  hence we do not add any $F$ term.

Using the equations of motion, disregarding  boundary terms and using the primary constraint $p_i^2=T^{2}t'^{2}$, one arrives at
\begin{equation}\label{NRPLs9}
	\dot G = \int\dif\sigma\ \frac{1}{T}\lambda\  \left(\partial_{i}\xi_{j}p_{j}p_{i} 
	- p_{i} E \partial_{i}\xi_0+p_{i}p_{i}\partial_{t}\xi_0+T^{2}t'x_{i}'\partial_{i}\xi_0\right).
\end{equation}
Since the functions $\xi_0$, $\xi_i$ do not depend on $p_i$ or the $\sigma$ derivatives of the several fields, the terms corresponding to all the powers of all these must be zero by themselves. One gets then the set of non-relativistic Killing equations  
\begin{eqnarray}
	\partial_{i}\xi_0&=&0,\label{NRPLs10}\\
	\frac{1}{2}\left(\partial_{i}\xi_{j}+\partial_{j}\xi_{i}\right)+\delta_{ij}\partial_{t}\xi_0&=&0,\label{NRPLs20}
\end{eqnarray}
where (\ref{NRPLs10}) comes from the second and fourth terms in (\ref{NRPLs9}), and  (\ref{NRPLs20}) 
from the quadratic terms in $p_i$.
For $i\neq j$ (\ref{NRPLs20}) implies 
\begin{equation}\label{killing}
	\partial_{i}\xi_{j}+\partial_{j}\xi_{i}=0,
\end{equation}
while for $i=j$,
\begin{equation}\label{killing2}
	\partial_{i}\xi_{i}+\partial_{t}\xi_0=0,\ i=1,\ldots, d.
\end{equation}
The general solution to (\ref{killing}) is 
\begin{equation}
	\xi_{i}(t,\vec{x})=a_{i}(t)+\omega_{ij}(t)x_{j}+c(t)x_{i},
\end{equation}
with 
\begin{equation}
	\omega_{ij}(t)=-\omega_{ji}(t).
\end{equation}
Substituting this into (\ref{killing2}) one gets then 
\begin{equation}
	c(t)+\partial_{t}\xi_0=0,
\end{equation}
with solution, 
\begin{equation}
	\xi_0(t,\vec{x})=-\int\text{d}t\ c(t)+\Psi\left(\vec{x}\right),
\end{equation}
but taking into account that $\partial_{i}\xi^0=0$
one gets
\begin{equation}
	\xi_0(t,\vec{x})=-\int\text{d}t\ c(t) + c_0.
\end{equation}

The transformations for the fields are
\begin{eqnarray}
	\delta t &=& -\int \text{d}u\ c(u) + c_0, \\
	\delta x_{i} &=& a_{i}(t)+\omega_{ij}(t)x_{j}+c(t)x_{i}. 
\end{eqnarray}
For $c=0$ one gets, as special cases, time translations, space translations, boosts and standard rotations, but also time-dependent rotations and higher order (in time) transformations of $x_{i}$. The transformations corresponding to  $c_0$, $a_i(t)$ and $\omega_ {ij}(t)$ constitute what is sometimes known as the Coriolis group \cite{Duval:1993pe}.

Particular cases of $c\neq0$ are 
\begin{itemize}
	\item[1.] $c(t)=c$, which yields $\delta x_{i}=cx_{i}$ and $\delta t=-ct$. These constitute the infinitesimal form of the scaling $x_{i}\to e^{c}x_{i}$ and $t\to e^{-c}t$.
	\item[2.] $c(t)=2\kappa t$, for which $\delta x_{i}=2\kappa tx_{i}$ and $\delta t=-\kappa t^{2}$.
\end{itemize}
If we set $c(t)=-\frac{\text{d}\beta(t)}{\text{d}t}$ and absorb $c_0$ into $\beta$, the above transformations can be rewritten as 
\begin{eqnarray}
	& & \delta t=\beta(t),\label{pldeltat} \\
	& & \delta x_{i}=a_{i}(t)+\omega_{ij}(t)x_{j}-\frac{\text{d}\beta(t)}{\text{d}t}x_{i},\label{pldeltaxi}
\end{eqnarray}
which contains $d+1+\frac{d(d-1)}{2}$ arbitrary functions of $t$. One can check that, if $B_i$ are the components of $\vec B$ given in (\ref{NRPLs1}),
\begin{equation}
	\delta B_{i}=\omega_{ij}(t)B_{j},
\end{equation}
and hence the action is invariant under these transformations, which is consistent with not considering any $F$ term in the generator.  Since the transformations that we have obtained depend on arbitrary  functions of $t$, one has in fact an infinite-dimensional algebra of transformations.

If we expand the arbitrary functions is powers of $t$,
\begin{eqnarray}
	a_i(t) &=& \sum_{n\geq -1} a_i^{(n)} t^{n+1},\\
	\omega_{ij}(t) &=& \sum_{n\geq 0} \omega_{ij}^{(n)} t^n,\\
	\beta(t) &=& \sum_{n\geq -1} \beta^{(n)} t^{n+1},	
\end{eqnarray}	
one can read off  from (\ref{pldeltat}) and (\ref{pldeltaxi}) the vector fields that generate the corresponding transformations
\begin{eqnarray}
	\hat{M}_i^{(n)} &=& t^{n+1}\partial_i,\ n\geq-1,\label{plMop}\\
	\hat{J}_{ij}^{(n)} &=& t^n (x_j\partial_i - x_i \partial_j),\ n\geq 0,\label{plJop}\\
	\hat{L}^{(n)} &=& t^{n+1}\partial_t - (n+1) t^n x_i \partial_i,\ n\geq-1.\label{plLop}	
\end{eqnarray}	
They generate the algebra
\begin{eqnarray}
	& & \left[\hat{M}_{i}^{\left(n\right)},\hat{M}_{j}^{\left(m\right)}\right]=0, \\
	& & \left[\hat{M}_{i}^{\left(n\right)},\hat{J}_{jk}^{\left(m\right)}\right]=\delta_{ik}\hat{M}_{j}^{\left(n+m\right)}-\delta_{ij}\hat{M}_{k}^{\left(n+m\right)}, \\
	& & \left[\hat{L}^{\left(n\right)},\hat{M}_{i}^{\left(m\right)}\right]= \left(n+m+2\right)\hat{M}_{i}^{\left(n+m\right)}, \label{cML}\\
	& & \left[\hat{L}^{\left(n\right)},\hat{L}^{\left(m\right)}\right]=\left(m-n\right)\hat{L}^{\left(n+m\right)}, \\
	& & \left[\hat{L}^{\left(n\right)},\hat{J}_{ij}^{\left(m\right)}\right]=m\hat{J}_{ij}^{\left(n+m\right)}, \\
	& & \left[\hat{J}_{ij}^{\left(n\right)},\hat{J}_{kl}^{\left(m\right)}\right]=\delta_{ik}\hat{J}_{jl}^{\left(n+m\right)}-\delta_{il}\hat{J}_{jk}^{\left(n+m\right)}+\delta_{jl}\hat{J}_{ik}^{\left(n+m\right)}-\delta_{jk}\hat{J}_{il}^{\left(n+m\right)}.
\end{eqnarray}
We have thus obtained an infinite dimensional extension  of the Galilean  algebra. 
Other algebras of the same type have been considered in the literature; see for instance \cite{Hosseiny:2009jj}\cite{Bagchi:2009my} and references therein.

The standard Galilean algebra is generated by
\begin{eqnarray}
	\hat{L}^{(-1)} &=& \partial_t\equiv \hat{H}\ \text{(time translations)},\\
	\hat{M}_i^{(-1)} &=& \partial_i\equiv \hat{P}_i\ \text{ (spatial translations)},\\
	\hat{M}_i^{(0)} &=& t\partial_i\equiv \hat{B}_i\ \text{(Galilean boosts)},\\ 
	\hat{J}_{ij}^{(0)} &=& x_j\partial_i - x_i \partial_j\equiv \hat{L}_{ij}\ \text{(spatial rotations)}. 
\end{eqnarray}
As in \cite{Bagchi:2009my}, one can try to extend this  algebra  by adding to these 
\begin{eqnarray}
	\hat{M}_i^{(1)} &=& t^2\partial_i\equiv \hat{K}_i\ \text{ (second order boosts)},\\
	\hat{L}^{(0)} &=& t\partial_t- x_i \partial_i\equiv \hat{D}\ \text{(dilatations)},\\
	\hat{L}^{(1)} &=& t^2\partial_t-2tx_i\partial_i\equiv \hat{K}\ \text{(special conformal transformations)}.
\end{eqnarray}
These generators are similar to those appearing in the Galilean conformal algebra\cite{Bagchi:2009my}\cite{Duval:2009vt}. Notice, however, that the relative sign in $\hat{D}$ and $\hat{K}$ between time and space directions is negative, while in the standard Galilean conformal algebra one has $\hat{D}=t\partial_t+ x_i \partial_i$, $\hat{K}= t^2\partial_t+2tx_i\partial_i$. Hence our extension of the Galilean algebra has dynamical exponent  $z=-1$, \textit{i.e.} space and time scale inversely one respect to the other. This can be seen directly from the action (\ref{LPLf}): the  scaling  that leaves the action invariant is $t\to e^{\lambda}t$, $\vec{x}\to e^{-\lambda}\vec{x}$.

The fact that the dynamical exponent is negative has as a  consequence that the numerical factor in (\ref{cML}) is always positive, except for 
$ \left[\hat{L}^{\left(-1\right)},\hat{M}_{i}^{\left(-1\right)}\right]=0$,
and this implies that any subalgebra containing any $\hat{M}_i^{(n)}$ and one $\hat{L}^{(m)}$ with $m\geq 1$ will contain also all of the   $\hat{M}_i^{(n)}$.
In particular,
\begin{equation}
	[\hat{L}^{(1)}, \hat{M}_i^{(1)}]	= 4 \hat{M}_i^{(2)},
\end{equation}	
and thus all of our extensions of the Galilean algebra  that contain special conformal transformations are infinite dimensional.

An infinite-dimensional extension of the Galilean algebra \cite{Henkel:1997zz}, called 
$z$-Galilean Conformal algebra  is discussed in  \cite{Hosseiny:2009jj} for the special case of a $2+1$ spacetime, with $1/z$
taking positive integer as well as half integer values. The algebra that we have obtained corresponds to a sub-algebra of an 
$z$-Galilean Conformal algebra one but with $z=-1$, and defined for any dimension of spacetime.

%
%
The generator of the symmetry transformations is
\begin{equation}
	G(\tau) = \int\dif\sigma \left(
	\left(
	a_i(t) + \omega_{ij}(t) x_j - \frac{\dif\beta(t)}{\dif t} x_i
	\right) p_i -\beta(t) E
	\right) 
	\label{plGen} 	
\end{equation}	
with all the variables ($x_i$, $t$, $p_i$ and $E$) functions of $\tau$ and $\sigma$.

Using the expansions in powers of $t$ yields the infinite set of charges
\begin{eqnarray}
	M_{i}^{(n)} &=&  \int\dif\sigma\  t^{n+1}p_i,\quad n\geq-1,\label{plMin}\\
	J_{ij}^{(n)} &=& \int\dif\sigma\  t^n (p_i x_j-p_j x_i),\quad n\geq 0,\label{plJijn}\\
	L^{(n)} &=& \int\dif\sigma\ \left( -t^{n+1}E -(n+1)t^n x_i p_i \right), \quad n\geq -1.\label{plLn}	
\end{eqnarray}

In \cite{Duval:2009vt} it is shown that for a massless non-relativistic particle the  generators of boost  and momenta are proportional. Equation (\ref{plMin}) for $n=0$ shows that in our case we have the analogous relation for the corresponding generator densities.

One can check that 
\begin{eqnarray}
	\dot{M}_{i}^{(n)} &=&  \int\dif\sigma\left(\mu t^{n+1} p_i\right)',\\
	\dot{J}_{ij}^{(n)}	&=& \int\dif\sigma\left( \mu t^n  (p_ix_j-p_j x_i)\right)',\\
	\dot{L}^{(n)} &=& \int\dif\sigma\left( - t^{n+1}(\mu E + T \lambda t')-(n+1)\mu t^n x_i  p_i\right)', 
\end{eqnarray}	   
and hence they are conserved if the boundary conditions (\ref{plBC2})(\ref{plBC3}) are used. This shows that, under the same boundary conditions, the boundary terms that we disregarded in the computation of $\dot G$ are actually zero.

The above conserved charges generate, under the Poisson bracket, an algebra which, except for an overall minus sign, coincides with the algebra of the vector fields which generate the transformations of $x_i$ and $t$. No non-central extensions appears because, as we remarked before, the Lagrangian density is exactly invariant under the transformations.

The analysis that we have developed for the particle limit of the string can be generalized to the case of the particle limit of  $p$-branes, and one ends up with the  following set of  vector fields for the symmetry transformations:
\begin{eqnarray}
	\hat{L}^{(n)} &=&  t^{n+1}\partial_t - \frac{1}{p} (n+1) t^n x_i \partial_i,\label{cbapbraneL}\\
	\hat{M}^{(n)}_i &=& t^{n+1}\partial_i,\label{cbapbraneM}\\
	\hat{J}_{ij}^{(n)} &=& -t^n (x_i\partial_j - x_j \partial_i) \equiv -t^n L_{ij},\label{cbapbraneJ}
\end{eqnarray}
the only difference with respect to the string ($p=1$) case being the $1/p$ factor in the spatial component of $\hat{L}^{(n)}$. It turns out that the algebra of these generators is exactly the same that in the string case, except for
\begin{equation}
	\left[ \hat{L}^{(n)}, \hat{M}_i^{(m)} \right] = \left( m+1+ \frac{1}{p}(n+1)  \right)M_i^{(n+m+1)},\label{cbapbraneLM}
\end{equation}
which reduces to (\ref{cML}) for $p=1$.  Hence we have obtained an infinite dimensional Galilean conformal  algebra with dynamical exponent $z=-p$.

\section{Spacetime symmetries of NR stringy limit of string}\label{stringlimit}

We consider now the most general generator of canonical point transformations for the non-relativistic string in $r, s, y_i$ variables, 
\begin{equation}\label{eqGrs}
	G=\int\dif\sigma\left( 
	p_i \xi_i(\vec y,r,s) + \pi_ r \xi_r(\vec y,r,s) +\pi_s \xi_s(\vec y,r,s)- F(\vec y,\vec y\,',r,r',s,s')
	\right),
\end{equation}
where the $F$ term has been added because the Lagrangian density is only pseudo-invariant under the stringy Galilei transformations.

Computing $\dot{G}$ one obtains, after some algebra and some integrations by parts,  
\begin{equation}\label{dotG}
	\begin{split}
		\dot G = & \int\dif\sigma \lambda (
		- T y_i' (\partial_j \xi_i y_j' + \partial_ r\xi_i r ' + \partial_s \xi_i s'\\
		+ &  \frac{1}{T}p_i p_j \partial_j \xi_i  - p_i \partial_r \xi_i r' + p_i \partial_s \xi_i s' \\
		+ & \pi_r  \partial_i \Phi_ r y_i' + 2\pi_r \partial_s\xi_r s' +\frac{1}{T} \pi_r p_i \partial_i \xi_r\\
		- & \pi_s \partial_i\xi_s y_i' - 2 \pi_s \partial_r\xi_s r' + \frac{1}{T} \pi_s p_i \partial_i\xi_s\\
		- & \frac{1}{T}p_i [F]_i + r' [F]_r - s' [F]_s 
		)\\
		- & \int\dif\sigma \mu (y_i' [F]_i +r'[F]_r+s' [F]_s),
	\end{split}
\end{equation}
where $[F]_x$ denotes the Lagrangian derivative of $F$ respect to $x$,
$
[F]_x = \partial_x F - \partial_\sigma(\partial_{x'}F)
$.
Demanding that this be zero imposes the following set of Killing equations
\begin{eqnarray}
	\partial_i\xi_r&=&0,\label{K1}\\
	\partial_i\xi_s&=&0,\label{K2}\\
	\partial_s\xi_r&=&0,\label{K3}\\
	\partial_r\xi_s&=&0,\label{K4}\\
	\partial_i \xi_j + \partial_j x_i &=&0,\label{K5}\\
	{[}F{]}_s &=& - T y_i' \partial_s\xi_i,\label{eqFs}\\
	{[}F{]}_r &=& Ty_i' \partial_r\xi_i,\label{eqFr}\\
	\frac{1}{T} [F]_i &=&  -\partial_r \xi_i r' +\partial_s \xi_i s',\label{eqFi} \\
	y_i' [F]_i + r' [F]_r + s' [F]_s &=& 0.\label{eqF3}
\end{eqnarray}
Equation (\ref{K1}) and (\ref{K2}) come from the terms  proportional to $\pi_r p_i$ and $\pi_s p_i$, respectively, and they also ensure the cancellation of the terms proportional to $\pi_r y_i'$ and $\pi_s y_i'$.
Likewise, the terms proportional to $\pi_r s'$ and $\pi_s r'$ lead to (\ref{K3}) and (\ref{K4}). Condition (\ref{K5}) comes from the terms proportional to $p_ip_j$ and $y_i'y_j'$.\footnote{Notice that these terms cannot be combined into the constraint $\Phi_0$ due to the relative sign.}
Finally, the last four equations are obtained  from the pieces proportional to $\lambda s'$, $\lambda r'$, $\lambda p_i$ and $\mu$.
Notice, however, that (\ref{eqF3}) is a linear combination of (\ref{eqFs}), (\ref{eqFr}) and (\ref{eqFi}), and hence can be disregarded.

From (\ref{K1})---(\ref{K4}) one has
\begin{eqnarray}
	\xi_r(\vec y,r,s) &=& f(r),\label{solPhir}\\
	\xi_s(\vec y,r,s) &=& g(s),\label{solPhis}
\end{eqnarray}
while the most general solution to (\ref{K5}) is
\begin{equation}\label{solxi}
	\xi_i = \Lambda_i(r,s) + \omega_{ij}(r,s) y_j,
\end{equation}
with the functions $\Lambda_i(r,s)$  and $\omega_{ij}$ to be determined, and with $\omega_{ij}$ skew-symmetric.

Given the structure of  (\ref{eqFs}) and (\ref{eqFr}), we make the ansatz that $\Lambda_i(r,s)$ and $\omega_{ij}(r,s)$ are, in fact, the sum of a function of $r$ and a function of $s$,
\begin{equation}
	\Lambda_i(r,s) =  B_i(r) +  \overline{B}_i(s),\quad \omega_{ij}(r,s) = \omega_{ij}(r) + \overline{\omega}_{ij}(s).
\end{equation}
Let us start with $\xi_i=B_i(r)$ and (\ref{eqFr}). One has that $\partial_r\xi_i=\partial_r B_i(r)$, and (\ref{eqFr}) is satisfied with
\begin{equation}
	F=T y_i'  B_i(r),
\end{equation}
without any condition on the function $B_i(r)$. Notice that, for the particular case that $B_i$ be a constant, $F$ becomes a derivative in $\sigma$ and hence its contribution to $G$ becomes a boundary term. Expanding $B_i(r)$ in powers of $r$ we get
\begin{equation}\label{Bxi}
	B_i(r) = \sum_{n\geq -1} b_i^{(n)} r^{n+1}
\end{equation}
with corresponding
\begin{equation}\label{Fxi}
	F=T y_i'\sum_{n\geq -1} b_i^{(n)} r^{(n+1)}.
\end{equation}

Due to the separation of $\Lambda$ as a sum of functions of $r$ and $s$, the solution to $(\ref{eqFs})$ will look exactly the same, with different parameters $\overline{b}_i^{(n)}$, and with  a global minus sign for the corresponding contribution to $F$. Thus, the solution to (\ref{eqFs}) and (\ref{eqFr}) of the form $\Lambda_i(r,s)$ will be (we combine the two constant contributions into a single one, $a_i=b_i^{(0)}+\overline{b}_i^{(0)}$, and disregard the corresponding total derivative contribution  $T (b_i^{(0)}-\overline{b}_i^{(0)}) y_i'$ in $F$)
\begin{eqnarray}
	\xi_i & =& a_i + \sum_{n\geq 0} \left(
	b_i^{(n)}  r^{n+1}
	+
	\overline{b}_i^{(n)}   s^{n+1}
	\right),\label{solxitot}\\
	F &=& T y_i'  \sum_{n\geq 0} \left( 
	b_i^{(n)} r^{n+1} - \overline{b}_i^{(n)} s^{n+1}
	\right).\label{solFtot}
\end{eqnarray}
It turns out that (\ref{solxitot})(\ref{solFtot}) also  solve (\ref{eqFi}).

Let us now consider the terms coming from $\omega_{ij}(r)$ and $\overline{\omega}_{ij}(s)$. Equations  (\ref{eqFs}) and (\ref{eqFr}) are satisfied with
\begin{equation}\label{Fw}
	F=T y_i' (\omega_{ij}(r)- \overline{\omega}_{ij}(s)) y_j,
\end{equation}
but, substituting into (\ref{eqFi}), one finds that, after canceling corresponding terms from both sides,
\begin{equation}
	\left(-\omega_{ij}(r)+\omega_{ji}(r)+\overline{\omega}_{ij}(s)- \overline{\omega}_{ji}(s)\right)y_j' =0,
\end{equation}
which only holds if $ \omega_{ij}(r)=\overline{\omega}_{ij}(s)=\omega_{ij}$, and then, from (\ref{Fw}), $F=0$.\footnote{As a direct check that only constant functions $\omega_{ij}$ (or $\overline{\omega}_{ij}$) yield symmetries of the action, one can consider
	$\delta y_i = \omega_{ij}(r) y_j$ and compute
	\begin{equation}
		\delta {\cal L}_{rs} = T y_i' y_j \dot\omega_{ij} - T \dot y_i y_j \omega_{ij}'.
	\end{equation}
	This cannot be written as $\partial_\tau G_\tau + \partial_\sigma G_\sigma$, unless $\dot\omega_{ij}=\omega_{ij}'=0$, in which case $\delta{\cal L}_{rs}=0$. For instance 
	\begin{equation}
		\partial_\tau (T y_i' y_j \omega_{ij}) + \partial_\sigma ( -T \dot y_i y_j \omega_{ij} ) =
		T y_i' y_j \dot\omega_{ij} - T \dot y_i y_j \omega_{ij}' + 2 T y_i' \dot y_j \omega_{ij},
	\end{equation}
	which has an extra term.}

Let us now return to the solutions (\ref{solPhir}) and (\ref{solPhis}). Since $f$ and $g$ are arbitrary, we can expand them in a power series  to obtain $f(r)=\sum_{n\geq -1}\alpha_n r^{n+1}$, $g(s)=\sum_{n\geq -1}\beta_n s^{n+1}$. Putting everything together  we write the final form of
the transformations as
\begin{eqnarray}
	\delta r &=& \sum_{n\geq -1}  \alpha_n r^{n+1},\label{deltarsl}\\
	\delta s &=&  \sum_{n\geq -1}  \beta_n s^{n+1},\label{deltassl}\\
	\delta y_i &=& a_i + \omega_{ij}y_j + \sum_{n\geq 0} \left(   b_i^{(n)} r^{n+1} + \overline{b}_i ^{(n)} s^{n+1} \right).\label{deltayisl}
\end{eqnarray}

These transformations have associated vector fields given by 
\begin{eqnarray}
	\hat{P}_i &=& \partial_i,\\
	\hat{L}_{ij} &=& y_j \partial_i - y_i \partial_j,\\
	\hat{L}^{(n)} &=& r^{n+1} \partial_r, \ n\geq -1,\\
	\hat{\overline{L}}^{(n)} &=& s^{n+1} \partial_s, \ n\geq -1,\\
	\hat{M}_i^{(n)} &=& r^{n+1} \partial_i,\ n\geq 0,\\
	\hat{\overline{M}}_i^{(n)} &=& s^{n+1} \partial_i,\ n\geq 0.
\end{eqnarray}
These generate two copies of the same algebra, which contains the Witt algebra, with relevant  commutators (the commutators involving $\hat P_i$ are those of $\hat M_i^{(n)}$ extended to $n=-1$)
\begin{eqnarray}
	\left[
	\hat{L}^{(n)}, \hat{L}^{(m)}
	\right] &=& -(n-m)\hat{L}^{(n+m)},\label{deW}\\
	\left[
	\hat{L}^{(n)}, \hat{M}_i^{(m)}
	\right] &=& (m+1)\hat{M}_i^{(n+m)},\label{commLnMni}\\
	\left[
	\hat{L}_{ij}, \hat{M}_k^{(n)}
	\right] &=&\delta_{ik} \hat{M}_j^{(n)}-\delta_{jk}\hat{M}_i^{(n)},\\
	\left[
	\hat{L}_{ij}, \hat{L}_{kl}
	\right] &=& \delta_{ik}\hat{L}_{jl}-\delta_{il}\hat{L}_{jk}+\delta_{jl}\hat{L}_{ik}-\delta_{jk}\hat{L}_{il}.
\end{eqnarray}
Notice that $\hat{P}_i$, $\hat{L}_{ij}$, $\hat L^{(-1)}$, $\hat{L}^{(0)}$ and $\hat{M}_i^{(0)}$ form a subalgebra, with associated transformations 
\begin{eqnarray}
	\delta r &=& 	\alpha_{-1} + \alpha_0 r,\\
	\delta y_i &=& a_i + \omega_{ij} y_j + b_i^{(0)}r,
\end{eqnarray}	
which correspond to translations and dilatations for the longitudinal light-cone variable $r$ and translations, rotations and Galilean boosts (in the longitudinal light-cone variable $r$) for the transverse variables $y_i$. One can add the operators $\hat{M}_i^{(1)}, \hat{M}_i^{(2)}, \ldots, \hat{M}_i^{(N)} $, $N\geq 1$, and still get a closed algebra, with transformations
\begin{eqnarray}
	\delta r &=& 	\alpha_{-1} + \alpha_0 r,\\
	\delta y_i &=& a_i + \omega_{ij} y_j + b_i^{(0)}r + b_i^{(1)}r^2 + b_i^{(2)}r^3 + \cdots + b_i^{(N)}r^{N+1} .
\end{eqnarray}

Because of (\ref{commLnMni}), adding to the above set any of the  $\hat{L}^{(n)}$, $n\geq 1$, forces the inclusion of all the $\hat{M}_i^{(n)}$.

In terms of the covariant variables $t=\hat{X}^0$, $z=\hat{X}^1$, the transformations are
\begin{eqnarray}
	\delta t &=& \frac{1}{2} g(t+z) + \frac{1}{2}f(t-z),\\
	\delta z &=& \frac{1}{2} g(t-z) - \frac{1}{2} f(t-z),\\
	\delta y_i &=& a_i + \omega_{ij}y_j + B_i(t-z) + \overline{B}_i(t+z).
\end{eqnarray}	
Expanding the arbitrary functions in powers of $t$ and $z$ one gets, to the lowest orders,
\begin{eqnarray}
	\delta t &=& \gamma_{-1} + \gamma_0 t + \overline{\gamma}_0 z + \gamma_1 t^2 + 2 \overline{\gamma}_1 t z + \gamma_1 z^2 + O(3),\\
	\delta z &=&  \overline{\gamma}_{-1} + \gamma_0 z + \overline{\gamma}_0 t + \overline{\gamma_1} t^2 + 2 \gamma_1 t z + \overline{\gamma}_1 z^2 + O(3),\\
	\delta y_i &=& a_i + \omega_{ij}y_j + c_i^{(0)}t +\overline{c}_i^{(0)}z + c_i^{(1)}t^2 + 2 \overline{c}_i^{(1)}tz + c_i^{(1)}z^2 +O(3),
\end{eqnarray}	
with
\begin{equation}
	\gamma_n = \frac{\beta_n + \alpha_n}{2},\ \ \overline{\gamma}_n = \frac{\beta_n-\alpha_n}{2},\  \
	c_i^{(n)} = \overline{b}_i^{(n)}+b_i^{(n)} ,  \ \ 	\overline{c}_i^{(n)} =\overline{b}_i^{(n)}- b_i^{(n)}. 
\end{equation}

If the expansion of the arbitrary function in powers of $r$ and $s$ is carried out in the canonical generator of the transformations one gets
\begin{equation}\label{finalG}
	\begin{split}
		G = &  \int\dif\sigma \left(
		p_i a_i + p_i \omega_{ij} y_j  + p_i \sum_{n\geq 0} \left(
		b_i^{(n)}  r^{n+1}
		+
		\overline{b}_i^{(n)}  s^{n+1}
		\right)\right.\\
		& + \pi_r \sum_{n\geq -1} \alpha_n r^{n+1} + \pi_s \sum_{n\geq -1} \beta_n s^{n+1} \\
		& \left.- T  y_i' \sum_{n\geq 0} \left(
		b_i^{(n)} r^{n+1} - \overline{b}_i^{(n)} s^{n+1}
		\right)
		\right).
	\end{split}
\end{equation}

From this one can read the charges
\begin{equation}\label{charges}
	\begin{split}
		P_i =& \int\dif\sigma\  p_i,\\
		L_{ij} =& \int\dif\sigma\  (y_i p_j - y_j p_i),\\
		L^{(n)} =&\int\dif\sigma\  \pi_r r^{n+1}, \ n\geq -1,\\
		\overline{L}^{(n)}  =& \int\dif\sigma\  \pi_s s^{n+1}, \ n\geq -1,\\
		M_{i}^{(n)}  =& \int\dif\sigma   r^{n+1}\left( p_i - Ty_i'\right), \ n\geq 0,\\
		\overline{M}_{i}^{(n)}  =& \int\dif\sigma  s^{n+1}\left( p_i  + T  y_i'\right), \ n\geq 0.
	\end{split}
\end{equation}

The computation of  $\dot G$, which has led to the charges (\ref{charges}), has been done performing several integrations by parts, which produce surface terms that we have disregarded. In order for the charges to be conserved, appropriate boundary conditions must be imposed so that those boundary terms are zero, and they can be obtained directly from the conservation of the charges. One has
\begin{eqnarray}
	\dot L_{ij} &=& \int\dif\sigma\left( y_i (T\lambda y_j' + \mu p_j) -  y_j (T\lambda y_i' + \mu p_i)\right)'        ,\label{dotLij}\\
	\dot P_i &=& \int\dif\sigma\left( T\lambda y_i' + \mu p_i\right)',\label{dotP}\\
	\dot L^{(n) }&=& \int\dif\sigma\left( -(\lambda-\mu) \pi_r r^{n+1}\right)',\label{dotLn}\\
	\dot M_i^{(n)}  &=& \int\dif\sigma\left((\lambda-\mu)r^{n+1}(Ty_i'-p_i)\right)',\label{dotMin}
\end{eqnarray}
and similar expressions  to (\ref{dotLn}) and (\ref{dotMin}) for the $\overline{L}^{(n)}$ and $\overline{M}^{(n)}_i$ charges, respectively.  
Hence, conditions (\ref{slBCa})(\ref{slBCb}) are sufficient for conservation of the charges in the open string case.

These  charges generate, up to possible non-central extensions which come from boundary terms, the same extension of  the Galilean  algebra obtained with  the vector fields associated to the transformation. 

The possibility of obtaining non-central extensions is due to the presence of a nonzero $F$ term in the canonical generator. 
Computing the Poisson brackets of the conserved charges, one obtains the same algebra (with a reversed sign), except for the fact that the brackets among $M_i^{(n)}$ (and among $\overline{M}_i^{(n)}$) are non-zero:
\begin{eqnarray}
	\left\{
	M_i^{(n)}, M_j^{(m)}
	\right\} 
	&=&  T \delta_{ij} \frac{n-m}{m+n+2} \int\dif\sigma \left(r^{m+n+2}\right)',\\
	\left\{
	P_i, M_j^{(m)}
	\right\} &=&  -T \delta_{ij} \int\dif\sigma \left(r^{m+1}\right)' .
\end{eqnarray}
(the second relation can be obtained from the first one with the extended index $n=-1$).
One can check that the new non-zero results do not spoil the closedness of the algebra.

\section{Discussion and outlook}\label{conclusions}

In this paper we have considered  different non-relativistic limits of relativistic extended objects. A $p$-brane in Minkowski space has $p+1$ target longitudinal directions, which   in the non-relativistic limit  could become large, and
therefore we can consider  $p+1$ different non-relativistic limits.  In the case of the string we have two limits that we call particle-limit , when only the temporal $X^0$ coordinate becomes large, and the stringy limit, when the two longitudinal directions are large.

In the first case  the string obtained does not vibrate, and physically it is a collection of non-relativistic free massless Galilean particles \cite{Duval:2009vt} whose energy density depends on the position of the particle in the string.  In the second case we have a string that vibrates. At quantum level,  if the spatial longitudinal directions are bounded, the theory is described by a conformal field theory with critical dimension 26
\cite{Gomis:2000bd}
\cite{Danielsson:2000gi}.

For both types of strings we have studied and solved the non-relativistic Killing equations. For the non-vibrating string we obtain symmetry transformations that close under an algebra 
\cite{Henkel:1997zz} which turns out to be an infinite dimensional extension of the Galilean  algebra 
with an exotic dynamical exponent $z=-1$.   In the vibrating case  we obtain a different infinite dimensional extension of the original  stringy Galilean algebra \cite{Brugues:2004an}. An infinite set of non-central extensions in the algebra of conserved charges is also obtained. If should be emphasized  that among the symmetries there is an infinite set of polynomial shift symmetries\cite{Griffin:2014bta}\cite{Horava:2016vkl}. The presence of polynomial shift symmetries has been noticed in Galileon theories\cite{Goon:2012dy}.

For future work it will be interesting to study in detail the action, dynamics and  symmetries of  all the $p+1$ non-relativistic limits of a $p$-brane.  For instance, in the case of a $2$-brane, if one considers the intermediate  NR stringy limit, one obtains a system that vibrates and  represents a continuum of non-relativistic vibrating strings
\cite{new}.

\section*{Acknowledgements}\addcontentsline{toc}{section}{Acknowledgements}
We acknowledge interesting discussions with Eric Bergshoeff, Roberto Casalbuoni, Josep Maria Pons and Paul Townsend.
{J.~Gomis} 
thanks the Galileo Galilei Institute for Theoretical Physics for the hospitality and the INFN for partial support during the completion of this work.
{C.~Batlle} is partially supported by  the Generalitat de Catalunya through project 2014 SGR 267 and by the Spanish government (MINECO/FEDER) under project CICYT DPI2015-69286-C3-2-R.
J.~Gomis has been supported in part by FPA2013-
46570-C2-1-P, 2014-SGR-104 (Generalitat de Catalunya) and Consolider CPAN and by
the Spanish goverment (MINECO/FEDER) under project MDM-2014-0369 of ICCUB (Unidad de Excelencia Mar\'\i a de
Maeztu).


\end{document}